\documentclass[fleqn,usenatbib]{mnras}

\usepackage[T1]{fontenc}

\DeclareRobustCommand{\VAN}[3]{#2}
\let\VANthebibliography\thebibliography
\def\thebibliography{\DeclareRobustCommand{\VAN}[3]{##3}\VANthebibliography}

\usepackage{graphicx}	
\usepackage{amsmath}	
\usepackage{amssymb}	
\usepackage{float}
\usepackage{longtable}
\usepackage{natbib}
\usepackage[toc,page]{appendix}
\usepackage{subfig}
\usepackage{multicol}
\usepackage{multirow}
\usepackage{xcolor}
\usepackage{lscape}
\usepackage{txfonts}
\usepackage{float}

\title[CHEOPS's hunt for exocomets]{CHEOPS's hunt for exocomets: photometric observations of 5 Vul}

\author[I. Rebollido et al.]{
Isabel Rebollido$^{1}$\thanks{E-mail: irebollido@stsci.edu},
Sebastian Zieba$^{2,3}$,
Daniela Iglesias$^{4}$,
Vincent Bourrier$^{5}$,
Flavien Kiefer$^{6}$
\newauthor
 and Alain Lecavelier Des Etangs$^{6}$
\\
$^{1}$Space Telescope Science Institute, Baltimore, MD 21218, USA \\
$^{2}$Max-Planck-Institut f\"ur Astronomie, K\"onigstuhl 17, D-69117 Heidelberg, Germany\\
$^{3}$Leiden Observatory, Leiden University, Niels Bohrweg 2, 2333CA Leiden, The Netherlands\\
$^{4}$School of Physics and Astronomy, Sir William Henry Bragg Building, University of Leeds, Leeds LS2 9JT, UK\\
$^{5}$Observatoire Astronomique de l' Universit\`e de Gen\`eve, Chemin Pegasi 51b, CH-1290 Versoix, Switzerland\\
$^{6}$Institut d' Astrophysique de Paris, Sorbonne Universit\`e, 98bis boulevard Arago, Paris, 75014, France\\
} 
\date{Accepted XXX. Received YYY; in original form ZZZ}
\pubyear{2022}
\begin{document}
\label{firstpage}
\pagerange{\pageref{firstpage}--\pageref{lastpage}}
\maketitle

\begin{abstract}
The presence of minor bodies in exoplanetary systems is in most cases inferred through infra-red excesses, with the exception of exocomets. Even if over 35 years have passed since the first detection of exocomets around $\beta$ Pic, only $\sim$ 25 systems are known to show evidence of evaporating bodies, and most of them have only been observed in spectroscopy. With the appearance of new high-precision photometric missions designed to search for exoplanets, such as {\it CHEOPS}, a new opportunity to detect exocomets is available. Combining data from {\it CHEOPS} and {\it TESS} we investigate the lightcurve of 5 Vul, an A-type star with detected variability in spectroscopy, to search for non periodic transits that could indicate the presence of dusty cometary tails in the system. While we did not find any evidence of minor bodies, the high precision of the data, along with the combination with previous spectroscopic results and models, allows for an estimation of the sizes and spatial distribution of the exocomets.

\end{abstract}




\section{Introduction}

Exocomets are still the only minor bodies we are able to observe in extrasolar planetary systems. 
However, they remain elusive, and since their first detection by \cite{Ferlet87} around the star $\beta$ Pictoris, only other $\sim$ 25 systems show evidence of the presence of such minor bodies \citep{strom20}.
The first evidences for exocomets were observed in spectroscopy, as (blue-)red-shifted variations in the Ca II K lines of several A-type stars, with a sample growing slowly throughout the years \citep[e.g.][]{Redfield08,Kiefer14a,Montgomery12,Rebollido20}. The variations observed spanned from few km/s to hundreds of km/s, and traced the gaseous tails of exocomets as they transited the star. They were found later in UV wavelengths, tracing other metallic elements \citep{vidalmadjar94,roberge00,grady18}
Due to the sporadic nature of exocometary events, their orbits are difficult to constrain, and only for the case of $\beta$ Pic we have estimations of the pericenter of the transiting comets, both through models \citep{BeustMorbidelli96,BeustMorbidelli00} and observations \citep{kennedy18}. 

Given comets in the solar system develop two tails, one composed of gas and another one composed of dust, it was predicted \citep{lecavelier99,lecavelier1999b} that photometric observations could also detect these bodies as individual (i.e., non periodic) transits, with a particular saw tooth shape due to the exponential decrease of material in the tail. 
While osbervations compatible with exocomets were detected with {\it Kepler} \citep{Boyajian16,Rappaport18, kennedy19}, the sensitivity and pointing constrains of the instrument did not allow for observations of the bright A-type stars where exocomets have been classicaly found using spectroscopy. 
Shortly after, the launch of {\it TESS} allowed monitoring of much brighter stars, leading to the detection of exocomets in photometry in the star $\beta$ Pic \citep{Zieba19,pavlenko2022} with a frequency high enough to make estimations about the size distribution of the minor bodies in the system \cite{lecavelier22}. To this date, there are no publication of simultaneously detected comets in spectroscopy and photometry around any star.

Aiming at expanding the sample of known spectroscopic exocomet host stars with photometric detections, we obtained {\it CHEOPS} Cycle 1 data of the star 5 Vulpecula, selected as at the time of the call for proposals it did not fall in the {\it TESS} observing windows. 
The structure of the paper is as follows: Section 2 describes the target and observations; Section 3 analyses the photometric upper limits and revises the published spectroscopic data; Section 4 offers an overview of the system and the potential discrepancy between observation strategies; and finally Section 5 summarises the work presented here.

\section{Target and observations}
\subsection{5 Vul}

5 Vul (HD 182919) is an A-type star with a detected debris disc \citep[Spitzer mid-IR data,][]{Morales09}. The most relevant data are summarized in Table \ref{tab:parameters}. \cite{Chen14} reported a L$_{*}$/L$_{IR}$ of 3.3x10$^{-5}$, and the presence of a double belt with temperatures of 295 K and 100 K for the inner and outer belts respectively.
The first evidence of a gaseous environment around 5 Vul was reported in \cite{Montgomery12}, where a FEB-like event was observed at $\sim$ 50 km/s, and further variations were also reported by \cite{Rebollido20}.

We selected 5 Vul as an optimal target for {\it CHEOPS} observations among other exocomet-host stars due to its proximity (less than 100 pc) and brightness (V$\sim$5.6). At the time when the call for proposals for {\it CHEOPS} Cycle 1 closed, 5 Vul was not expected to be observed by the primary mission of {\it TESS} as it fell in the CCD gap, unlike other exocomet-host stars. It was also not observed by {\it K2} due to its height relative to the ecliptic and brightness (V=5.59). Posterior changes in {\it TESS} schedule permitted observations of the target, that are also included in this work.

\begin{table*}
\caption{Stellar parameters for 5 Vul}
\label{tab:parameters}

\begin{tabular}{cccccccccccc}
\hline
\hline
Name &RA(2000.0)& DEC(2000.0) & SpT  & $v_{rad}$ & $T_{eff}$  & log$g$ &  $v \sin(i)$ & Distance & V  & Age   \\

 & & & & (km s$^{-1}$) & (K) & [cgs] &  (km s$^{-1}$) & (pc) & mag & Myr  \\
\hline
5 Vul  & 19:26:13.25 & +20:05:51.8 & A0V & -24.3 $\pm$ 1.4 & 10460 $\pm$ 80 & 4.47 $\pm$ 0.10 & 154 & 71.98  & 5.59 & 198\\
\end{tabular}\\

{\bf Notes:} Distance and coordinates were obtanied from Gaia \citep{Gaia,GaiaDR2}, age from \cite{Chen14}, and vrad, Teff, log g and vsini from \cite{Rebollido20}. V mag is from Simbad. 

\end{table*}

\subsection{Observations and data reduction}
\label{sect:obs}
\subsubsection{CHEOPS}

\begin{figure}
\centering
	\includegraphics[width=0.73\columnwidth]{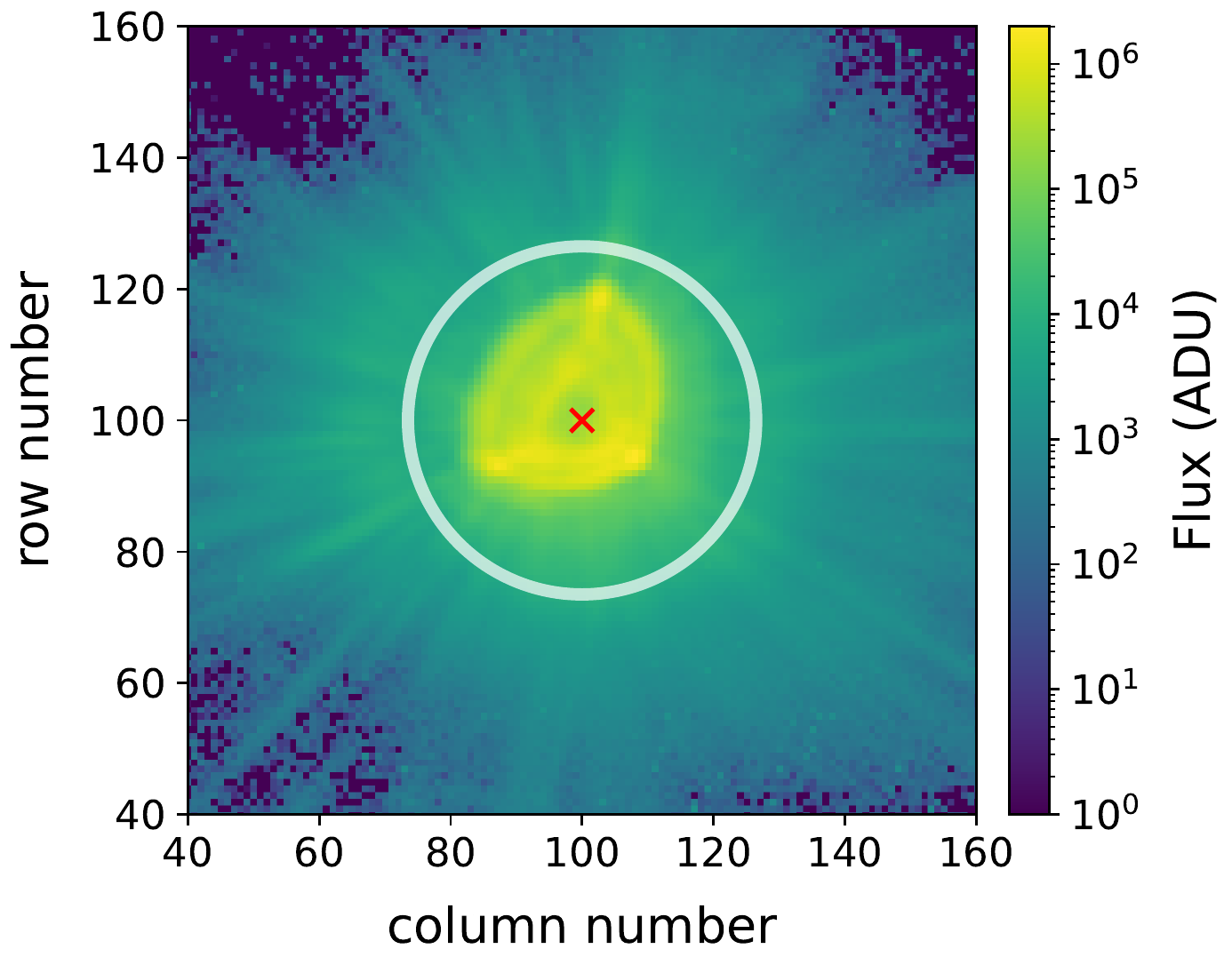}
    \caption{A {\it CHEOPS} exposure showing the typical PSF shape of the instrument, including the centroid of the star as determined by the DRP marked with a red cross and the OPTIMAL aperture being shown with a white circle.}
    \label{fig:cheops_psf}
\end{figure}

\begin{figure}
	\includegraphics[width=\columnwidth]{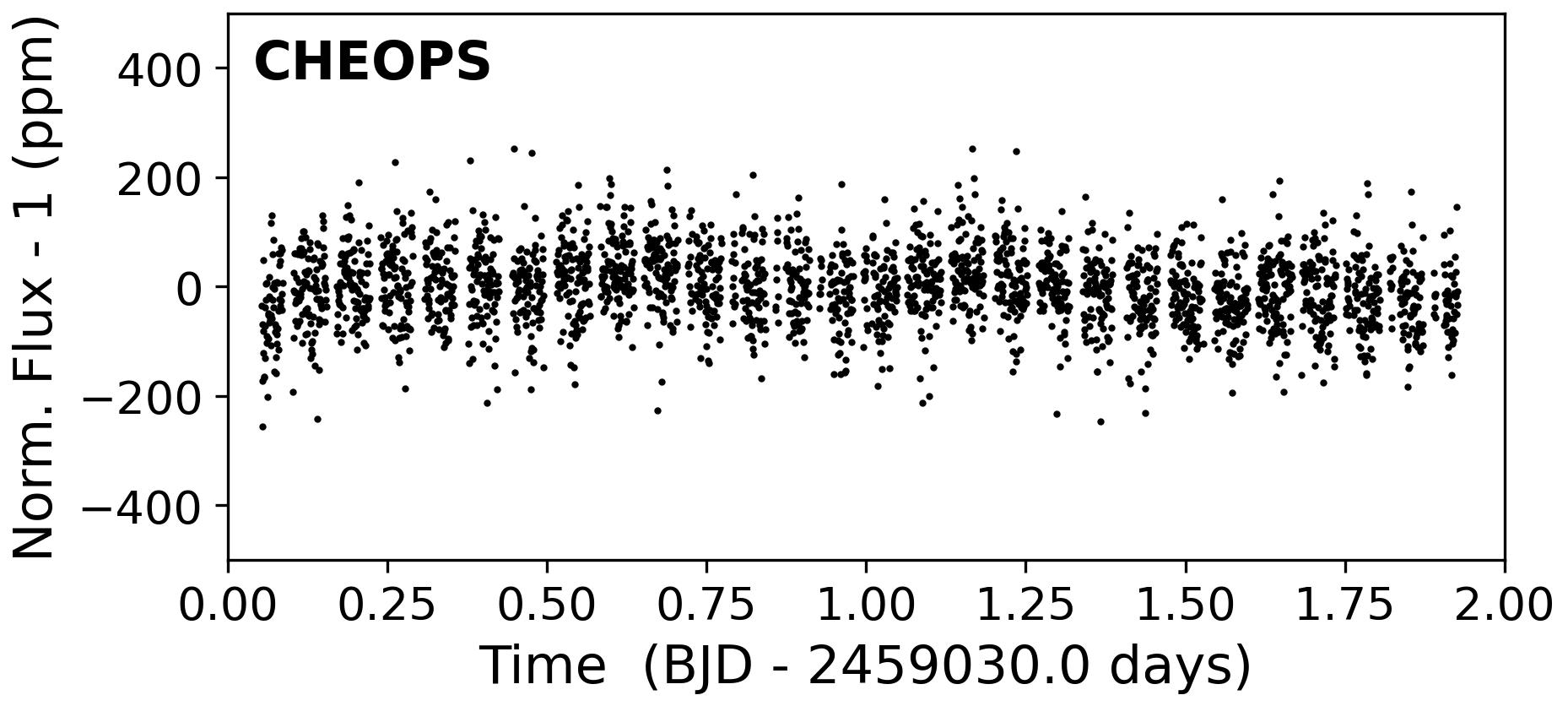}
 	\includegraphics[width=\columnwidth]{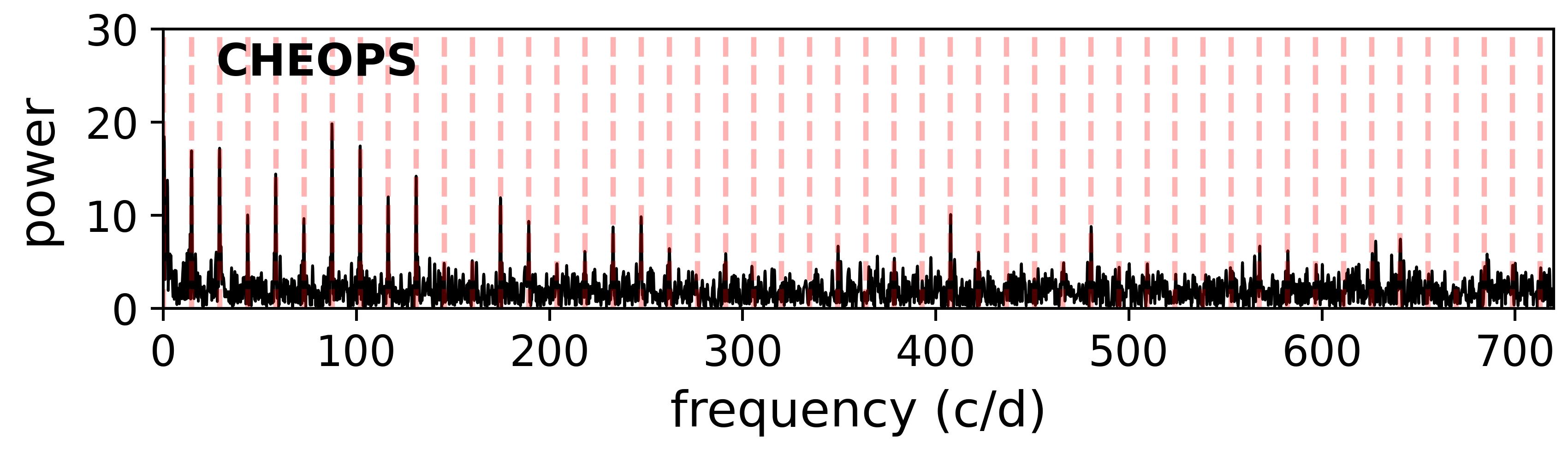}
    \caption{Top: Full {\it CHEOPS} lightcurve. Normalized and outliers removed. Bottom: Periodogram corresponding to {\it CHEOPS} data. Vertical red dashed lines shows peaks at multiples of 14.5 cpd ($\sim$ 100 minutes), corresponding to the orbital breaks of the satellite.}
    \label{fig:cheops_lc}
\end{figure}

The {\it CHEOPS} \citep{Benz21} data were taken as part of program CH\_PR210021 (``Hunting for exocomets transiting the young naked-eye star 5 Vulpeculae'', PI Rebollido) between June 29$^{th}$ and July 01$^{st}$ 2020 (see Table \ref{tab:dates}). Due to the objective of observing a non-periodic transit, the observations were targeted to be non-interrupted, and led to one visit over a duration of 44.96 hours. Data were processed with the latest version of the {\it CHEOPS} automatic Data Reduction Pipeline (DRP v13.1.0). The pipeline corrects the raw images (bias subtraction, gain conversion, flat fielding, dark correction and non-linearity) and then performs aperture photometry. An example for a CHEOPS exposure can be found in Fig. \ref{fig:cheops_psf}. A detailed description of DRP can be found in \citet{Hoyer2020}. The pipeline outputs light curves with differently sized apertures: R = 25.0 pixels (DEFAULT), R = 22.5 pixels (RINF), R = 30.0 pixels (RSUP), and R = 26.5 (OPTIMAL). The latter maximizes the signal-to-noise ratio (SNR) of the photometry by maximizing the flux coming from the target while minimizing the flux from background stars. This SNR calculation, performed by the DRP, uses the Gaia catalogue and the PSF shape of CHEOPS. The OPTIMAL aperture has a resulting point-to-point root-mean-square (rms) of 63.9 ppm, which is the lowest of all considered apertures. We therefore chose the OPTIMAL aperture for our analysis. Figure \ref{fig:cheops_lc}, top pannel, shows the {\it CHEOPS} light curve used in this analysis. Flagged observations (which might indicate cosmic ray events or crossing of the South Atlantic Anomaly) and outliers greater than 4$\sigma$ with respect to the median have been removed. These make up approximately 4\% of the total observations. The corresponding periodogram is shown in the bottom pannel of Fig. \ref{fig:cheops_lc}. The vertical dashed lines indicate the multiples of 14.5 cpd ($\sim$ 100 min), corresponding to the orbital breaks of the satellite. No other frequencies show significant peaks.

\begin{table*}
\captionsetup{justification=centering}
\centering
\caption{Observations with {\it TESS} and {\it CHEOPS} used in this work.}
\label{tab:dates}
\renewcommand{\arraystretch}{1.2}
\begin{tabular}{c|c|c|c}\hline\hline
Observation & Start Date (UTC)      & End Date (UTC)        & Exp. time (s) \\\hline
TESSS14    & 2019-Jul-18 20:35:54 & 2019-Aug-14 16:59:19 & 120 \\
TESSS40    & 2021-Jun-25 03:46:55 & 2021-Jul-23 08:35:25 & 120 \\
TESSS54    & 2022-Jul-09 09:41:08 & 2022-Aug-04 15:11:01 & 120 \\
CHEOPS     & 2020-Jun-29 13:16:11 & 2020-Jul-01 10:13:44 & 44
\end{tabular}
\renewcommand{\arraystretch}{1.2}
\end{table*}

\subsubsection{TESS}

\begin{figure*}
	\includegraphics[scale=0.4]{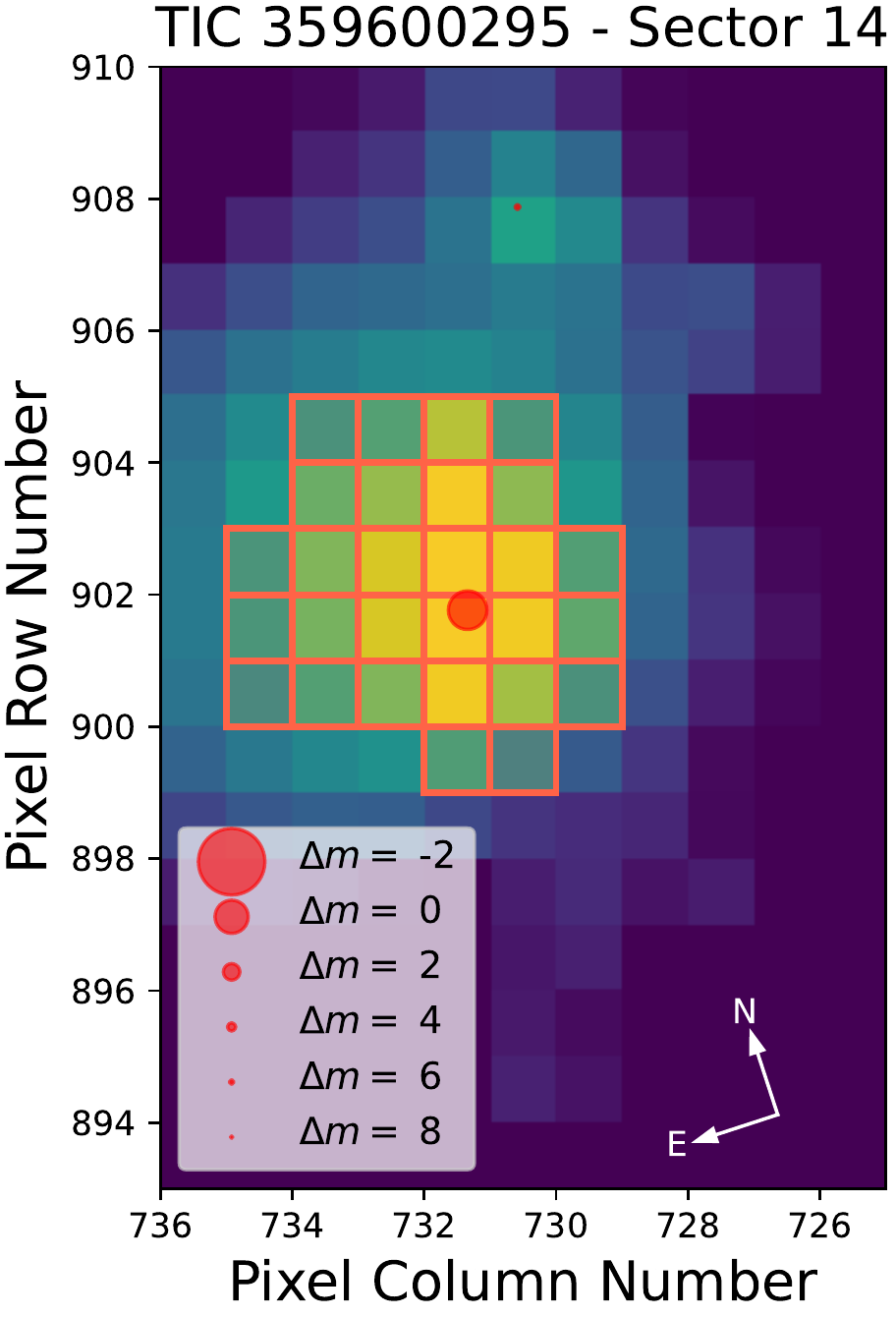}
	\includegraphics[scale=0.4]{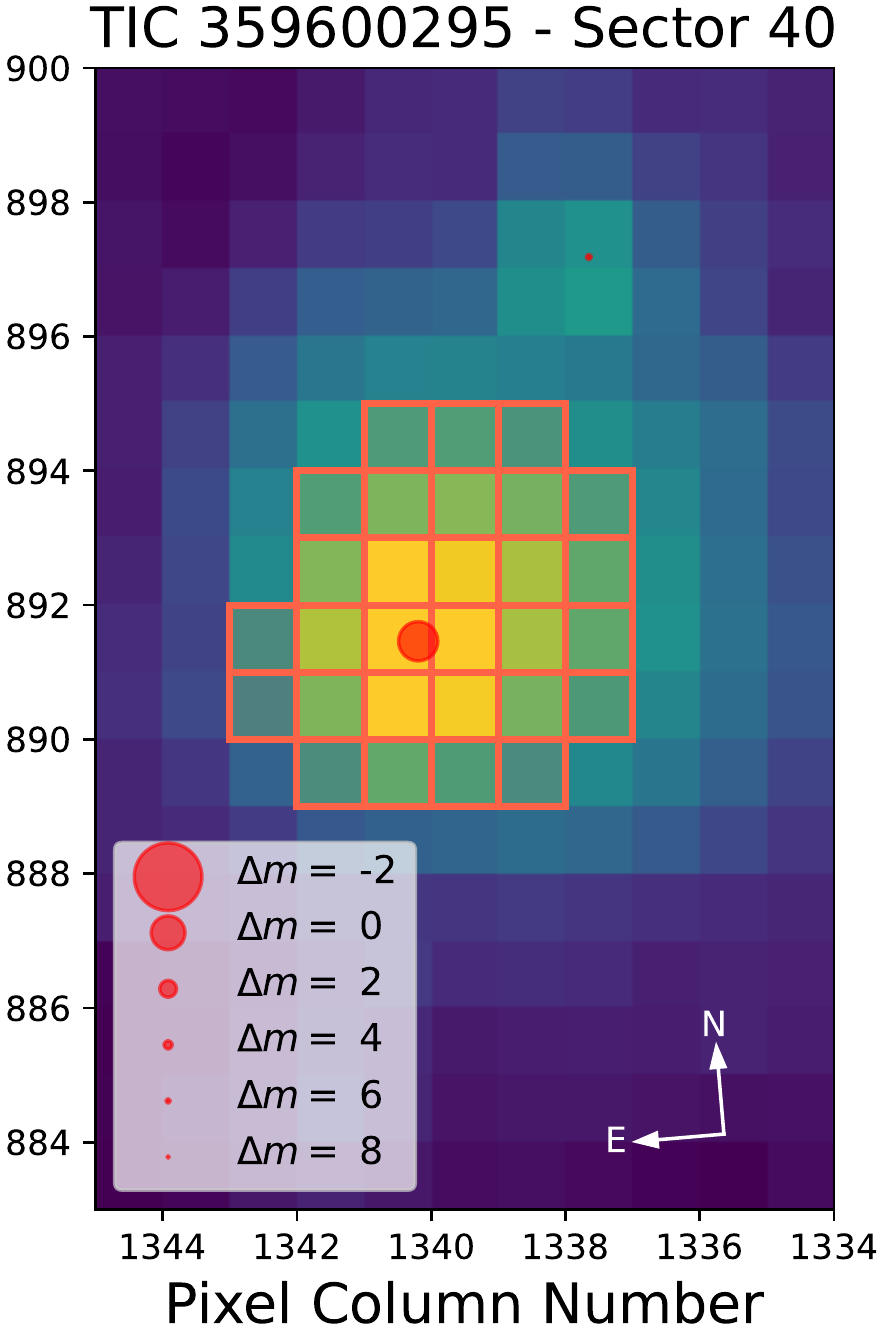}
 	\includegraphics[scale=0.4]{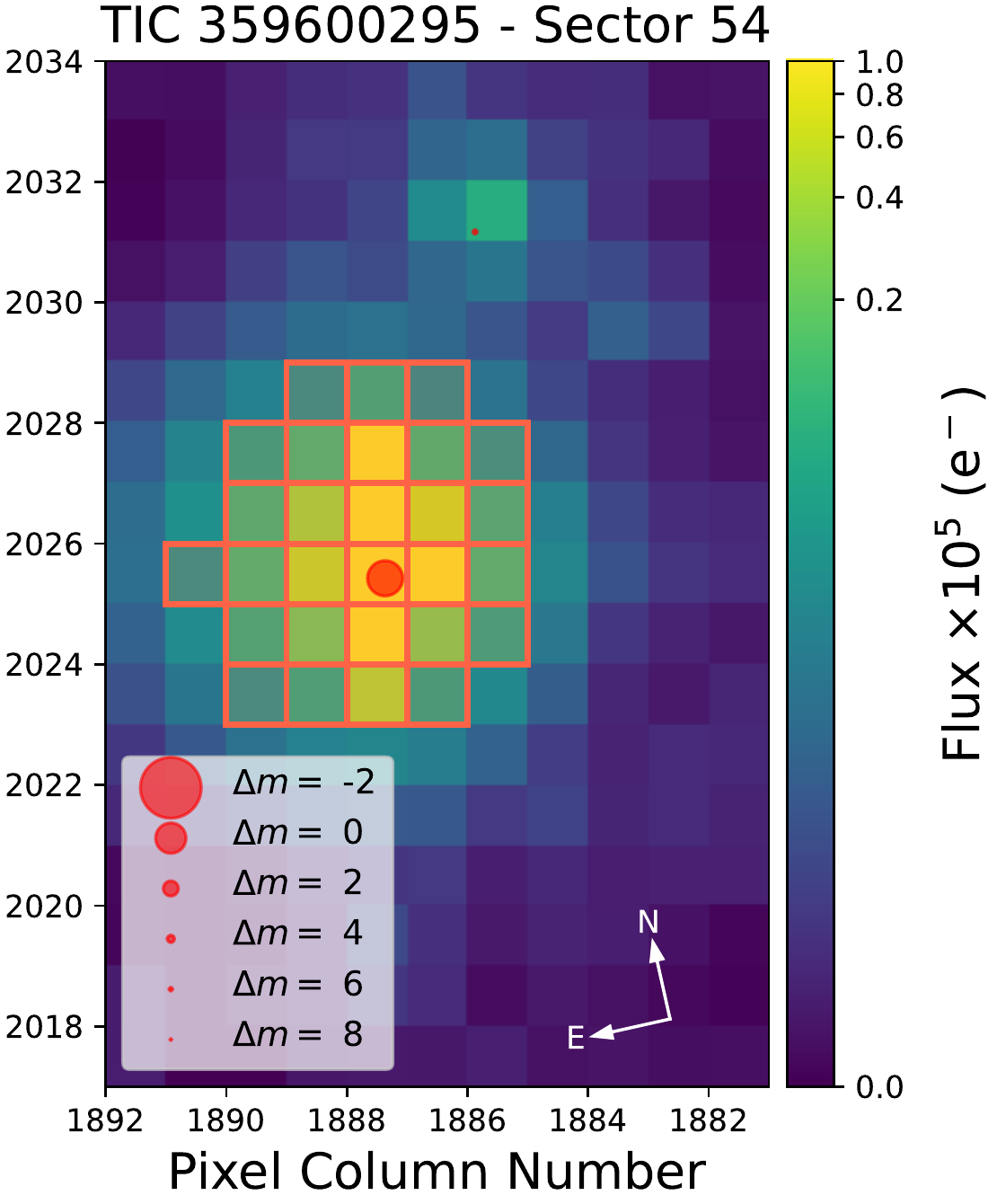}
    \caption{ {\it TESS} target pixel files (TPFs) for 5 Vul in Sector 14, 40 and 54. The pixels shaded in red indicate the aperture used by the SPOC pipeline. There are no stars within the aperture of 5 Vul with a Gaia magnitude difference smaller than 8, meaning ackground stars do not significantly contribute to the measured flux of 5 Vul. Plot made using \texttt{tpfplotter} \citep{Aller2020}.}
    \label{fig:tess_tpf}
\end{figure*}

\begin{figure}
	\includegraphics[width=\columnwidth]{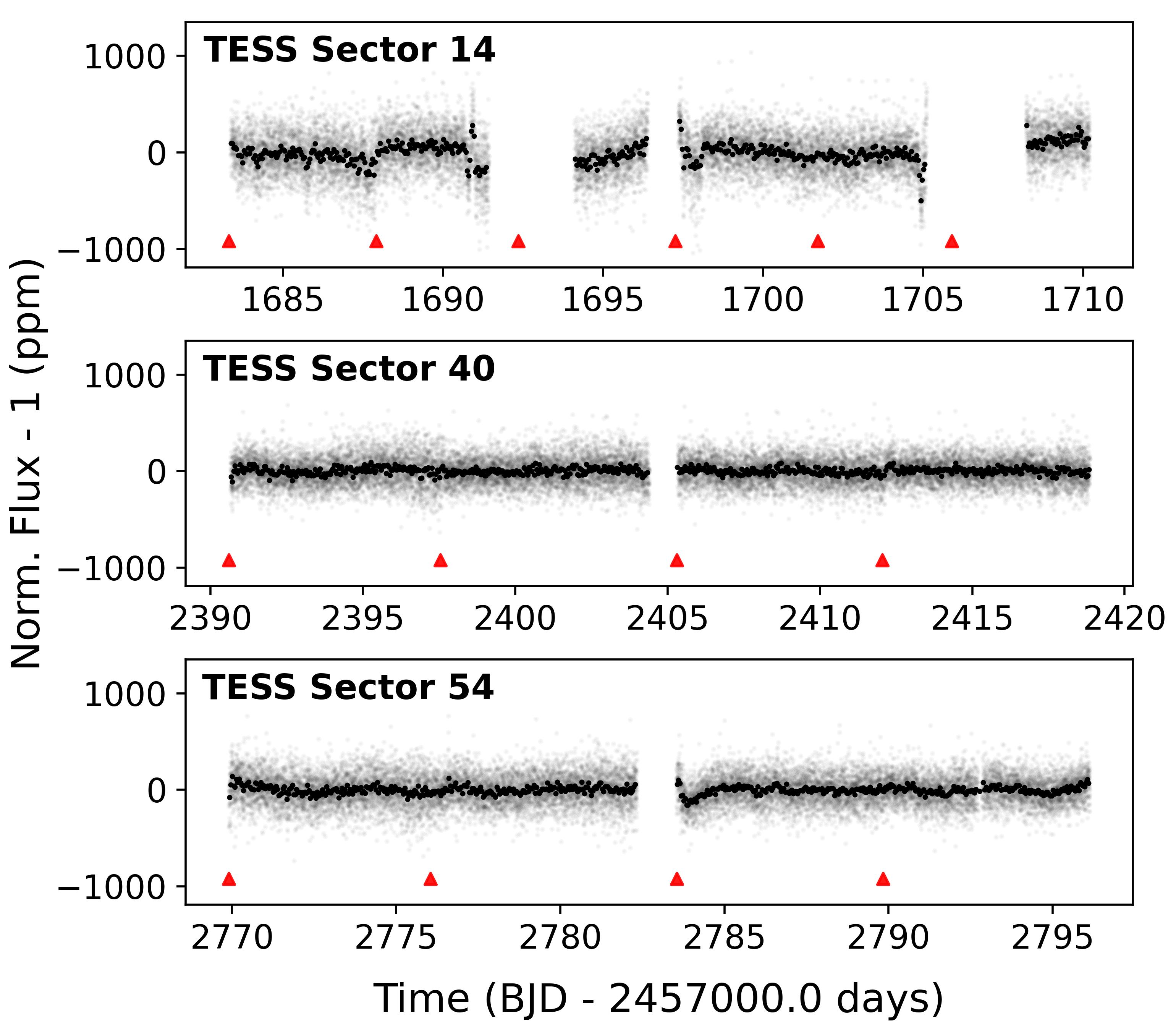}
    \caption{Full {\it TESS} lightcurve. Telescope momentum dumps are marked with red arrows. The 2-minute cadence TESS data (shown in light gray) have been binned to timescales of an hour for better visualisation.}
    \label{fig:tess_lc}
\end{figure}

5 Vul (TIC 359600295) was observed by {\it TESS} \citep{Ricker2015} in Sector 14 from 2019 July 18 to 2019 August 14, in Sector 40 from 2021 June 25 to 2021 July 23 and in Sector 54 from 2022 July 09 to 2022 August 04 at a 2 minute cadence (see Table \ref{tab:dates}).
Data were processed by the {\it TESS} Science Processing Operations Center (SPOC) pipeline \citep{Jenkins2016} and accessed using the python package \texttt{lightkurve} \citep{Lightkurve2018} which downloads the data from the Mikulski Archive for Space Telescopes (MAST) archive\footnote{\url{https://archive.stsci.edu/missions-and-data/tess}}. Target pixel files are shown in Fig. \ref{fig:tess_tpf}.
For this analysis, we used the Pre-Search Data Conditioning Simple Aperture Photometry \citep[PDCSAP;][]{Smith2012, Stumpe2012, Stumpe2014} light curves. In contrast to the Simple Aperture Photometry (SAP) light curves, the PDCSAP data are corrected for instrumental systematic effects and show considerably less scatter and variability caused by instrumental events like momentum dumps.
The PDCSAP light curves were flagged by the SPOC pipeline for bad data which mark anomalies like instrumental issues or cosmic ray events. We removed any {\it TESS} exposure in our dataset with a non-zero “quality” flag. For Sector 14, scattered light from the Earth was saturating the part of the detector where 5 Vul hit on silicon\footnote{For more information, see the Data Release Note of Sector 14: \url{https://archive.stsci.edu/missions/tess/doc/tess_drn/tess_sector_14_drn19_v02.pdf}}. We excluded these times which occurred in the last quarter of each orbit (see Fig. \ref{fig:tess_lc} around BTJD\footnote{BTJD (Barycentric TESS Julian Date) = BJD - 2457000.0 days} 1691-1694 and BTJD 1705-1708). The breaks of approximately one day in the middle of each Sector, are related to the data downlink of {\it TESS} when it reaches its perigee. Figure \ref{fig:tess_lc} shows that any significant changes in flux occur at the beginning or at the end of an {\it TESS} orbit or during  momentum dumps and are therefore not caused by the star itself. In total, we removed approximately 7\% of the {\it TESS} data mostly due to saturation of Camera 1 in Sector 14. Figure \ref{fig:tess_tpf} shows the pointing of {\it TESS} in Sector 14, 40 and 54, showing that there are no close, bright stars nearby which could bias our flux measurements.

\begin{figure}
	\includegraphics[width=\columnwidth]{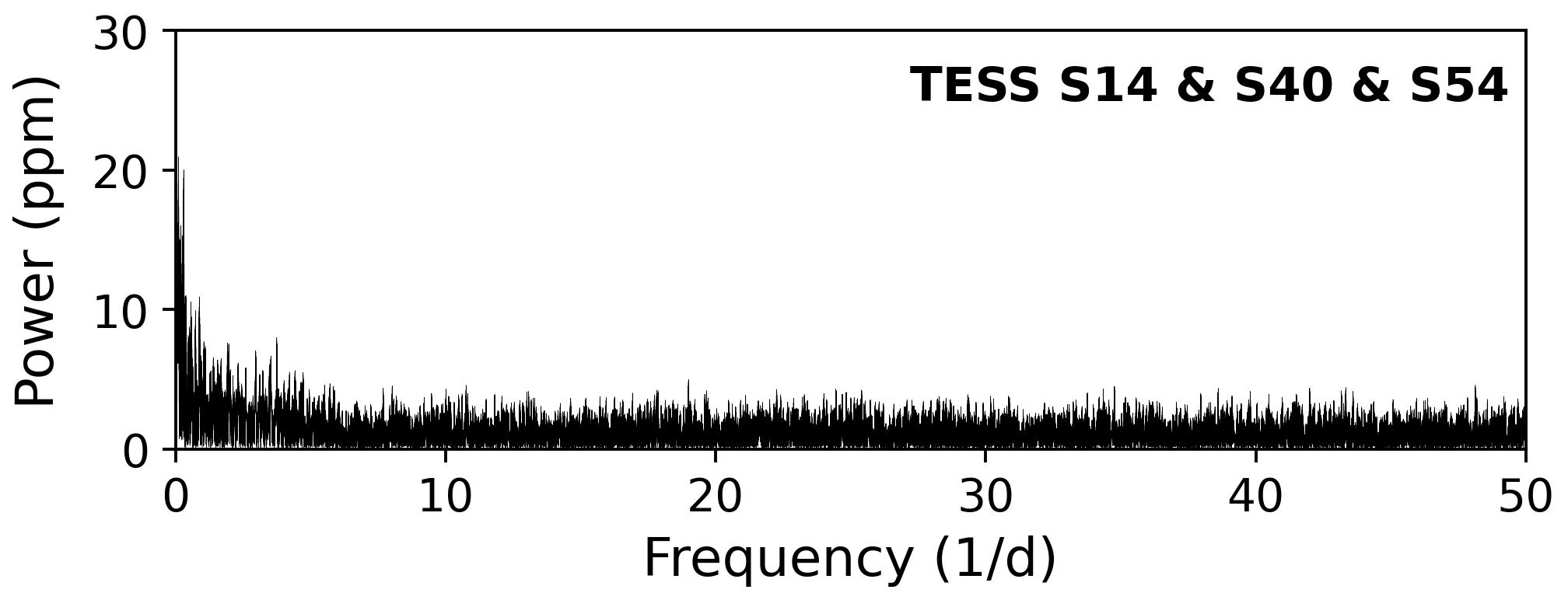}
    \caption{Lomb–Scargle periodogram of the full {\it TESS} lightcurve}
    \label{fig:tess_fft}
\end{figure}

The {\it TESS} data do not show any significant periodic signals which could be attributed to stellar variability (see Figure \ref{fig:tess_fft}). Systematic signals at very short periods, such as the orbit of TESS around Earth, with a period of 14 days can be hinted from Fig. \ref{fig:tess_lc}.

\section{Analysis of Results}


There is no detection of flux fluctuations indicative of the presence of exocometary activity in the light curve of 5 Vul or any other transits with a significative depth (>5$\sigma$) after the analysis reported in Sec. \ref{sect:obs}. However, we can estimate comet sizes for a given trajectory based on the upper limits.  

\subsection{Detectability}

Exocomets have been previously detected with {\it Kepler} \citep[formally K2, ][]{Boyajian16,Rappaport18} and {\it TESS} \citep{Zieba19}. While those missions were not designed for this particular science case, both have provided interesting results. The first detection of exocomets around a star with a spectral type different than A was in photometry, using K2 data \citep{Rappaport18}, and the detection of photometric exocomets around $\beta$ Pic, the only star with exocomet detection using two different techniques, used {\it TESS} data \citep{Zieba19}. 

When compared to those missions, {\it CHEOPS} provides very similar capabilities. The cadence of {\it CHEOPS} observations is shorter (1 min), but comparable to {\it TESS} and {\it Kepler} (2 min). The photometric precision, however, is much better, with an estimated 10 ppm for a V=6 star against $\sim$ 50 ppm for {\it TESS} \citep{Ricker2015} and while a number of bright targets were observed with Kepler \citep[e.g.][]{Guzik16},  the magnitude of 5 Vul exceeded Kepler's nominal mission, and it was never observed.
The response of the detectors is very unlikely to be responsible for detection rates either, since it is very similar to {\it TESS} and practically identical to {\it Kepler} \footnote{More information about the {\it CHEOPS} bandpass and its comparison with previous missions in \url{https://www.cosmos.esa.int/web/cheops/performances-bandpass}}.

Therefore, the non-detection in this target is most likely related to the lack of exocometary transits at the time of observations, as explained in Sect. \ref{sect:spectra}, and not to the instrumental capabilities.

\subsection{Maximum exocomet size}
Given that there is previous evidence of exocomets in the system, we explore the range of sizes and periastrons that we are sensitive to.  
We propose two different approaches for the size estimation, and test them for the different detection limits of both observatories.

\subsubsection{Estimation based on Hill spheres}
\label{sect:hill_size}

Assuming exocomets have very eccentric orbits \citep{BeustMorbidelli00}, the more volatile materials evaporate as they come closer to the central star, developing a coma composed by the evaporating gas and the dust dragged by it. 
If we consider all the material in the coma to be optically thick and gravitationally bound to the nucleus, we can follow \cite{Boyajian16} approximation, and estimate the maximum exocomet nucleus size we would be able to detect given the signal to noise ratio (SNR) of our lightcurve. 

The depth of the transit, $\tau$, is directly related to the surface of the star covered by the comet, and for two spherical bodies can be expressed as a function of the ratio of their radius, $r$ and $R_*$ for the comet and star respectively,
$$
    \tau = \left( \frac{r}{R_*} \right) ^2
$$

Given our SNR, we would not be able to detect transits smaller than 5.85$\cdot$10$^{-3}$\% (58 ppm) for {\it TESS} and 1.38$\cdot$10$^{-3}$\% (13 ppm) for {\it CHEOPS} (see Figs. \ref{fig:cheops_lc},\ref{fig:tess_lc}).


To retrieve the size of the comet associated to the clump, we can take into account the definition of Hill radius as:
$$
R_{Hill} = a(1-e)~\left( \frac{M_{comet}}{3M_{*}}\right) ^{1/3}
$$

Given we are estimating all our material is optically thick, we can assume $\tau \sim R_{Hill}$. 
The minimum periastron values are limited by our cadence (1 min for {\it CHEOPS} and 2 min for {\it TESS}) following Kepler's third law, and are consistent with the estimations for exocometary orbits for $\beta$ Pic \citep{lecavelier22}. The obtained mass for the cometary nucleus can then be converted to radius considering a typical density for a comet of $\mathrm{0.5~g/cm^3}$ \citep{britt06}. 

\begin{figure}
	\includegraphics[width=\columnwidth]{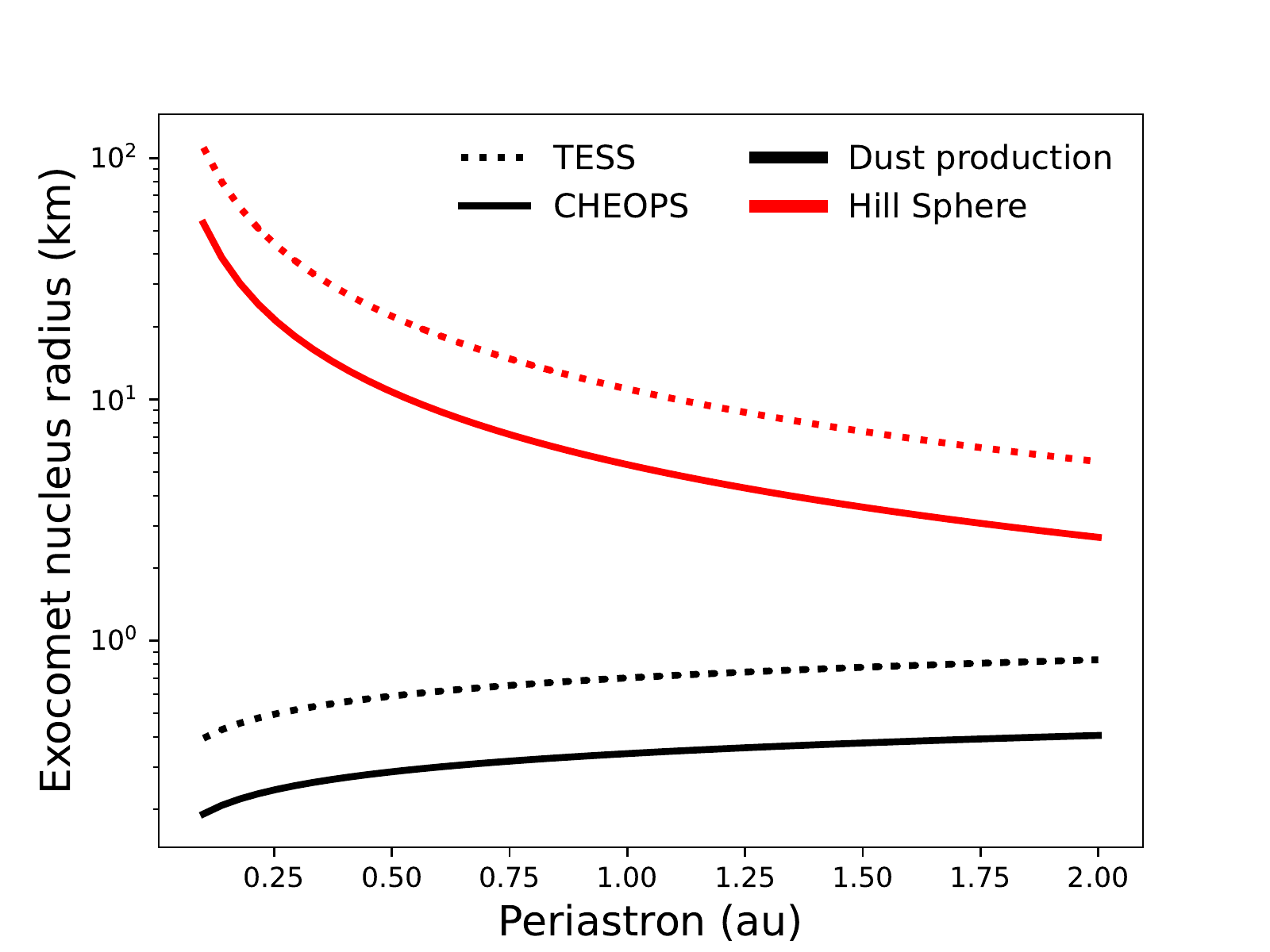}
    \caption{Minimum exocomet nucleus size estimates accessible with both {\it CHEOPS} (solid line) and {\it TESS} (dotted line) observations as a function of periastron distance. Red lines show size estimations for a spherical exocometary body with all the opaque material gravitationally bound to the central body (Hill sphere). Black lines show size estimations based on dust production rates.}
    
    \label{fig:exocomet_estimation}
\end{figure}

Fig. \ref{fig:exocomet_estimation} shows in red the minimum size of the exocometary nucleus for an A0V star \citep[2.193 R$_\odot$ and 2.18 M$_\odot$][]{PecautMamajek13}. From this calculation we can estimate that any body that transited the star during the observations must have had a nucleus smaller than $\sim$ 5.2 km for {\it CHEOPS} and $\sim$ 10.7 km for {\it TESS} at a distance of 1 AU. The fact that the increase in distance allows to trace smaller bodies is based on our first assumption of all material released in the evaporation process is contained in the Hill sphere and optically thick, which might not be a realistic approximation. Actually, assuming a similar composition throughout the system, the further from the star the exocomet is, the less material we are able to extract from the comet due to the inverse squared dependence of stellar flux with distance. Therefore, in the next section we explore a size estimation that contains evaporation models. 

\subsubsection{Estimations based on dust production rates}

\label{sect:dust production rates}
Following the calculation made by \cite{lecavelier22}, we can estimate the corresponding minimum exocomet dust production rate that we would be able to detect with our lightcurves. 

The typical absorption depth ($AD$) in a light curve that is caused by the transit of an exocomet is given by \citet{lecavelier22}:
$$
AD=5\cdot 10^{-5} \left( \frac{\dot{M_{1 {\rm au}}}}{10^5 \, {\rm kg/s}}\right)
\left(\frac{q}{1\, {\rm au}}\right)^{-1/2}
\left( \frac{M_{\star}}{{\rm M}_{\sun}}\right),
$$
where $\dot{M_{1 {\rm au}}}$ is the comet dust production rate taken at 1\,au from the star, 
$q$ is the comet orbit periastron distance, and $M_\star$ the mass of the star. 
Therefore a detection limit in absorption depth can be translated into a corresponding maximum dust production rate.
For a stellar mass $M_{\star}$=2.18\,M$_\odot$ \citep{PecautMamajek13}, we obtain 
$$
\dot{M_{1 {\rm au}}} = 1.2\cdot 10^4 \, {\rm kg/s}
\left(\frac{AD}{13\, {\rm ppm}}\right)
\left(\frac{q}{1\, {\rm au}}\right)^{1/2}
$$
Recalling that the Hale-Bopp comet had a dust prodution rate in the order of $2\cdot 10^6$\,kg/s at 1\,au from the Sun, we see that the {\it CHEOPS} observations are very sensitive to small comets.

Similarly as done by \cite{lecavelier22}, the dust production rate can be converted into a corresponding radius of the cometary nucleus using the relationship:
$$
\dot{M_{1 {\rm au}}} =  2\cdot 10^6 \, {\rm kg/s} 
\left(\frac{R}{30\, {\rm km}}\right)^2
\left(
\frac{L_\star}{{\rm L}_{\sun}}\right)
,
$$
where $R$ is the radius of the comet nucleus and $L_\star$ the stellar luminosity, and we are considering a typical cometary radius and dust production similar to Hale-Bopp \citep{Jewitt99,Fernandez99}.
With a luminosity of about 40\,L$_{\sun}$ for the A0 star 5~Vul \citep{yoon2010}, we finally have 
$$
R= 0.36\, {\rm km} 
\left(\frac{AD}{13\, {\rm ppm}}\right)^{1/2}
\left(\frac{q}{1\, {\rm au}}\right)^{1/4}.
$$

When taking into account the evaporation models, it appears that the presented {\it CHEOPS} and {\it TESS} observations with no exocomet photometric transit detection allow us to exclude the transit of very small bodies (0.3 km for {\it CHEOPS} and 0.7 km for {\it TESS}) at 1\,au over the observation period.

\section{discussion}

The detection of exocomets using photometric data has been restricted to a few systems so far. Contrary to what is observed in spectroscopy, where most comets are found around A-type stars {\citep[e.g.][]{Rappaport18, kennedy19}}, the detections in photometry do not seem to be restricted to a certain spectral type.  
In the following we discuss the properties of 5 Vulpeculae, and what might be the origin of the discrepancy of the spectroscopic and photometric data. 

\subsection{Disc and planets}

A faint debris disc is located in the environment of 5 Vul. \cite{Chen14} report significant excess for wavelengths longer than 24 $\mu$m corresponding to a faint and likely not very massive disc, specially when compared to other debris discs around A-type stars. They fit two components to the disc, with two different blackbody (BB) temperatures, being the most massive the colder component (10$^{-5}$ M$_{\earth}$), at a distance of 34\,au and with a BB temperature of $\sim$ 100 K. 
More recently, \cite{Musso21} reported a detection of the disc in band L, and fitted a single BB at $\sim$23 au, with a temperature of $\sim$ 180 K. The main goal of that paper was to look for planetary companions, and they also report 5$\sigma$ mass limits for all the investigated stars, including 5 Vul. However, the precision achieved can only estimate an upper limit for our system of >20 M$_J$ at distances shorter than 50 au, twice the mass of $\beta$ Pic b. Their analysis of the self-stirring mechanisms \citep[see Fig. A4 in][]{Musso21} indicate that the disc is not large enough to produce the observed dust, and possibly a perturber (planet, companion, binary star) is affecting the system dynamics. 
\cite{Matthews18} also reported no planetary companions larger than 8 M$_{J}$ at distances larger than 10\,au  based on SPHERE observations in H2 and H3.
Despite having hot Ca {\sc ii} gas detected, located very close to the star (see Fig. \ref{fig:phot_vs_spec}) and a known debris disc, there is no detection of cold gas in the outer regions of the system \citep[see e.g.][ and references therein for an overview of CO gas around A-type stars]{marino20}. \cite{Rebollido22} report an upper mass limit for the dust and CO content based on ALMA observations of $\sim$ 10$^{-3}$ and 10$^{-6}$ M$_{\earth}$ respectively, consistent with previous models \citep{kral17}.
The age of 5 Vul, estimated around 200 Myr \citep{Chen14} could potentially explain the low fractional luminosity and the lack of CO gas due to a decrease in dynamical activity as the system settles.

\subsection{Spectroscopic counterpart in 5 Vul}
\label{sect:spectra}

\begin{figure}
	\includegraphics[width=\columnwidth]{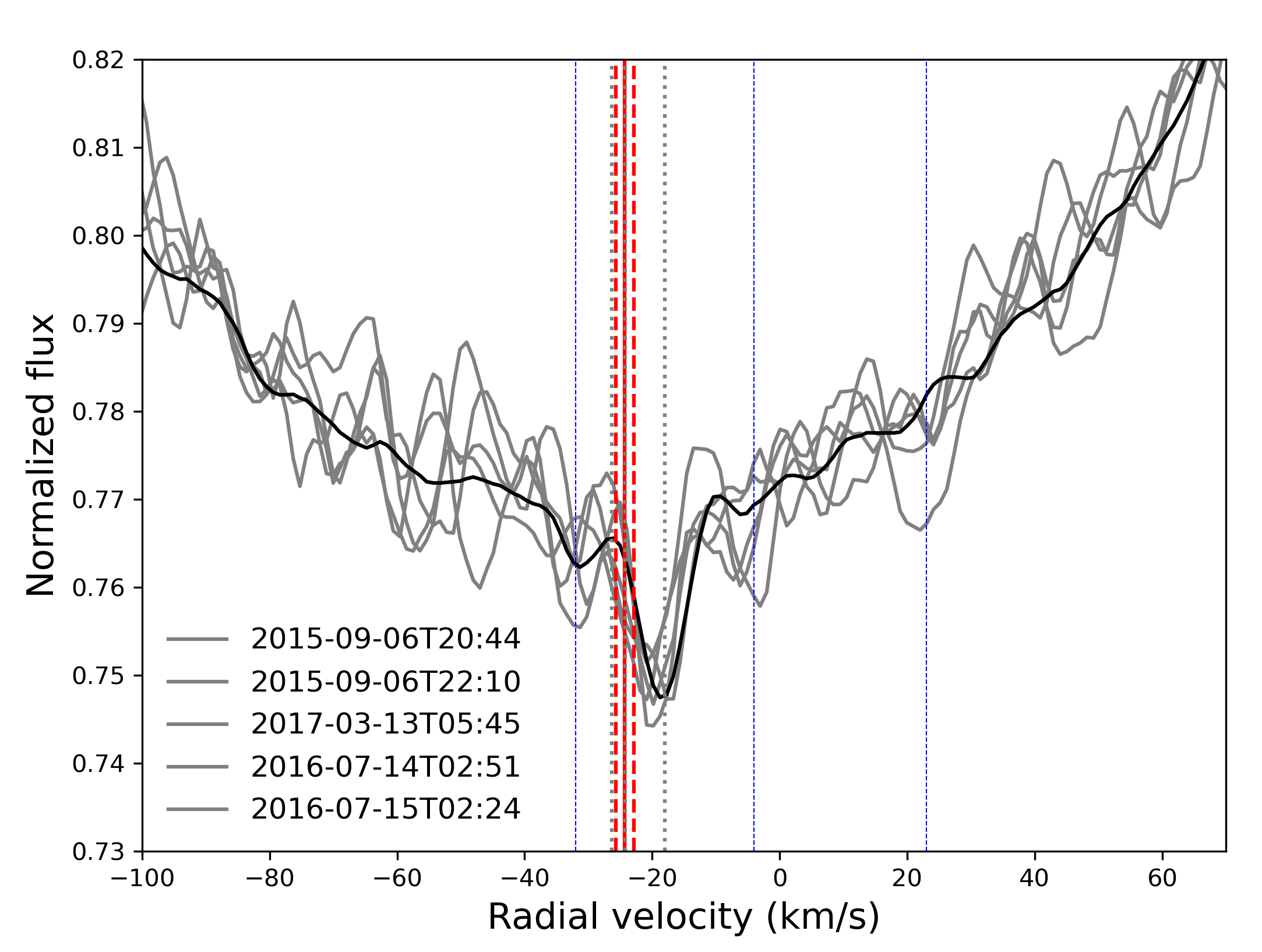}
    \caption{Spectroscopic evidence of exocomets in 5 Vul. Variations are seen at -3.5, 23 and -31 km/s, indicated by the blue vertical lines. Red vertical lines show the radial velocity of the object (-24.3 $\pm$ 1.4 km/s), and grey vertical lines show the radial velocity of the ISM in the line of sight (-18.05, -24.21, -26.30 km/s). }
    
    \label{fig:spectra}
\end{figure}

The investigation of the 5 Vul spectra has revealed the presence of exocomets, reported in \cite{Montgomery12, Rebollido20}. 
We show in Fig. \ref{fig:spectra} spectra previously published in \cite{Rebollido20} and publicly available online \footnote{ESO archive and \url{https://doi.org/10.26093/cds/vizier.36390011}}. The exocometary events were detected within -5 and 60 km/s in both cases, with one extra tentative detection located at -35 km/s. The poor time coverage does not allow for a follow up of the events. 
Moreover, both papers report variability of the more stable component \citep[at $\sim$-20. km/s consistent with the radial velocity of the star, but also with the G interstellar cloud][]{Redfield08}, suggesting at least a partially circumstellar origin and variations in the amount of circumstellar gas. 

There are 39 spectra in \cite{Rebollido20} spanning 22 nights, which overall show variable absorptions both blue and red-shifted in approximately 20\% of the observations. This is consistent with a frequency of one detected variation every 4.2 days. 
\cite{Montgomery12} show one exocometary red-shifted absorption in 4 spectra, and small variations in the EW measurements of the stable feature that seem gradual with time \citep[see Fig. 3 and Table 2 of][]{Montgomery12}, again consistent with variations in ~20\% of the observations.




\subsection{Photometric comets vs. Spectroscopic comets}

\begin{figure*}
	\includegraphics[width=1.5\columnwidth]{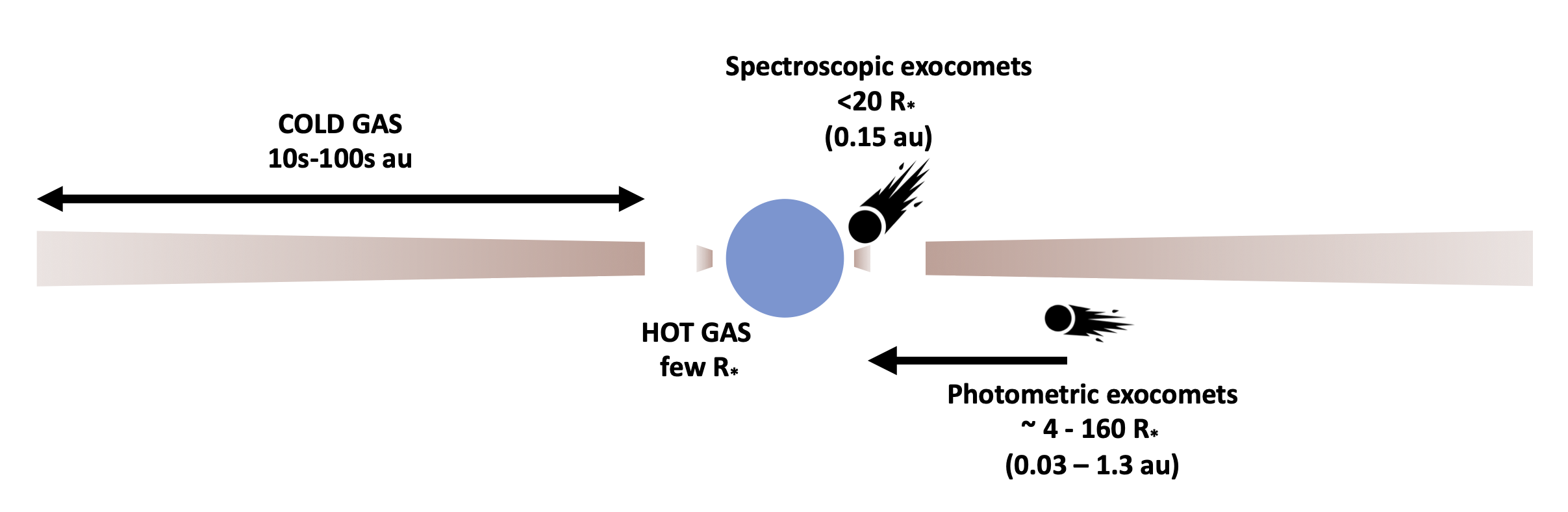}
    \caption{Estimated distribution of gas and exocometary bodies in a typical exocometary system. }
    
    \label{fig:phot_vs_spec}
\end{figure*}

The only star where exocomets have been found both in spectroscopy and photometry is $\beta$ Pic \citep[][]{Ferlet87, Kiefer14b, Zieba19}
. The high frequency observed in spectroscopy, of around one exocometary absorption observed every hour to six hours, surpasses greatly any other exocomet host system \citep{Kiefer14b}. However, the frequency of the photometric exocomets is much lower, with 30~events detected in 156 days through 4 different {\it TESS} sectors \citep{lecavelier22}. 
The key to explain this discrepancy could be in the expected stellocentric distances traced by both techniques: while spectroscopic exocomets are expected to be located very close to the star \citep[below 20R$_*$, i.e. <0.15 au,][]{Kiefer14b,kennedy18}, photometric ones are estimated at longer distances \citep[$\sim$ 4-160 R$_*$, i.e. 0.03 to 1.3 au; with an estimated average distance of 0.18 au, ][]{lecavelier22}. The question remains of whether different exocomet populations could be feeding different gas populations in the disc, i.e. gas detected in emission, much more extended \citep[e.g.][]{Marino16,matra19b,Rebollido22,Kospal13,moor17,moor19} and hot gas detected in absorption, potentially released by exocomets and located closer to the star \citep[e.g.][]{hobbs85,Ferlet87,Montgomery12,Iglesias2018,iglesias19,Rebollido20}. A diagram for the exocomet vs. gas location is shown Fig \ref{fig:phot_vs_spec}.

However, while the number of exocomets detected in spectroscopy remains larger than those detected in photometry, population D from \cite{Kiefer14b} is more likely to be in better agreement with the orbital ranges of the photometric populations. 


If we translate these exocomet ratios to the observed frequency around 5 Vul in spectroscopy, we would be expecting one photometric transit every 87.36 days, which would be hard to cover with current instrumentation and/or space missions.


\section{Summary}

Observations performed with {\it CHEOPS} and {\it TESS} of the star 5 Vulpecula show no evidence of exocometary activity in the lightcurve, despite the exocomet detection in spectroscopy around this star. In this work, we provided an estimation of the sizes and spatial location of the exocomets and a possible explanation for the non detection of exocometary transits via photometry. 

The sporadic nature of exocomets makes it difficult to trace them, as their orbits are almost impossible to constrain. The few efforts to detect the dust counterpart of the exocomets observed in spectroscopy have only been successful for $\beta$ Pic. This is not surprising, given the high frequency of exocomets, so far much higher than any other. Even for $\beta$ Pic, only a handful of exocomets have been observed in photometry, contrasting the thousands of events that are reported in spectroscopy. This could potentially be related to the distance at which the exocomets evaporate/sublimate. The estimated distances for Ca {\sc ii} production (where most exocomets have been detected) is just a few stellar radii, much closer to the estimated distances for the photometric observations. This could indicate we are probing two different groups of exocomets: star-grazing comets that sublimate refractory materials (i.e., calcium), and comets at larger orbits, where the cloud of dust is better sustained.

\cite{lecavelier99} estimated dozens of exocometary detections in photometry for large surveys with tens of thousands of stars in the worst case scenario. However, the results from large missions like {\it Kepler} and {\it TESS} show otherwise, with very few detections so far \citep[e.g.][]{ansdell19,Zieba19,Boyajian16}. Given that as of today a large enough sample of stellar ages and spectral types have been observed, the number of detectable exocomets might have been overestimated based on the activity around $\beta$ Pic.

\section*{Acknowledgements}

We thank the Lorentz Centre for facilitating the workshop "ExoComets: Understanding the Composition of Planetary Building Blocks" during 2019 May 13–17. The authors are grateful to the staff of the Lorentz Center and to the scientific organisers of the workshop.
{\it CHEOPS} is an ESA mission in partnership with Switzerland with important contributions to the payload and the ground segment from Austria, Belgium, France, Germany, Hungary, Italy, Portugal, Spain, Sweden, and the United Kingdom.
This paper also includes data collected by the {\it TESS} mission, obtained from the MAST data archive at the Space Telescope Science Institute (STScI).
I.R. thanks Grant Kennedy for the insightful conversations during the Spirit of Lyot 2022 conference in Leiden.
A.L and F.K. acknowledge support from the CNES (Centre national d’\'etudes spatiales, France).
This work made use of \texttt{tpfplotter} by J. Lillo-Box (publicly available in www.github.com/jlillo/tpfplotter), which also made use of the python packages \texttt{astropy}, \texttt{lightkurve}, \texttt{matplotlib} and \texttt{numpy}. 

\section*{Data Availability}

Data are available in the {\it CHEOPS} and {\it TESS} archives: \newline https://cheops-archive.astro.unige.ch/
\newline https://archive.stsci.edu/missions-and-data/tess




\bibliographystyle{mnras}
\bibliography{5vul_bib} 








\bsp	
\label{lastpage}
\end{document}